\documentclass[english,11pt]{revtex4}
\usepackage{graphics}
\usepackage{siunitx}
\usepackage{babel}
\begin{document}
\title{ Absorption spectroscopy of xenon and ethylene-noble gas mixtures at high pressure: Towards Bose-Einstein condensation of vacuum ultraviolet photons}
\author{Christian Wahl, Rudolf Brausemann, Julian Schmitt, Frank Vewinger, Stavros Christopoulos, Martin Weitz\footnote{Correspondence to: weitz@uni-bonn.de}
}                     
\affiliation{Institut f{\"u}r Angewandte Physik, Universit{\"a}t Bonn\\
Wegelerstra{\ss}e 8, 53115 Bonn, Germany}

\noindent

\begin{abstract}
Bose-Einstein condensation is a phenomenon well known for material particles as cold atomic gases, and this concept has in recent years been extended to photons confined in microscopic optical cavities. 
Essential for the operation of such a photon condensate is a thermalization mechanism that conserves the average particle number, as in the visible spectral regime can be realized by subsequent absorption re-emission processes in dye molecules. 
Here we report on the status of an experimental effort aiming at the extension of the concept of Bose-Einstein condensation of photons towards the vacuum ultraviolet spectral regime, with gases at high pressure conditions serving as a thermalization medium for the photon gas.
We have recorded absorption spectra of xenon gas at up to \SI{30}{\bar} gas pressure of the $5p^6 - 5p^56s$ transition with a wavelength close to \SI{147}{\nano\meter}.
Moreover, spectra of ethylene noble gas mixtures between \SIrange[range-phrase = { and }]{155}{180}{\nano\meter} wavelength are reported.
\end{abstract}
\maketitle
\section{Introduction}
\label{intro}\noindent
For bosonic particles with a non-vanishing rest mass Bose-Einstein condensation to a macroscopically occupied ground state minimizes the free energy when cooled to very low temperature given a sufficient particle density \cite{einstein1914quantentheorie, cornell2002nobel, ketterle2002nobel, anderson1995observation}.
In distinct contrast, Bose-Einstein condensation usually does not occur for photons \cite{Huang}.
For the perhaps most widely discussed photon gas, blackbody radiation \cite{planck1901law,bose1924plancks}, photons disappear in the system walls when cooled to low temperature instead of exhibiting condensation to the ground mode.
This is what is expressed by the common statement of the chemical potential vanishing for the photon case.
Early theory work has proposed photon Bose-Einstein condensation in Compton scattering of X-rays \cite{zel1969bose}, and Chiao et al. proposed a photon fluid in a nonlinear cavity \cite{chiao2000bogoliubov, bolda2001dissipative}.
More recently, Bose-Einstein condensation of exciton-polaritons, which are mixed states of matter and light in the strongly coupled regime, has been observed experimentally in several groups \cite{kasprzak2006bose, deng2010exciton, balili2007bose}.
Collisions of the material part of the polaritons here drives the system into or near thermal equilibrium.

Our group has in 2010 observed Bose-Einstein condensation of photons in a dye-solution filled microcavity \cite{klaersbose2010, klaers2011bose, Schmitt}, a result confirmed more recently by Marelic and Nyman \cite{marelic2015experimental}.
A cavity with very small mirror spacing here imprints a low-frequency cutoff for photons in the visible spectral range, yielding a range of allowed photon frequencies well above the thermal energy in frequency units.
The confinement results in the photon gas becoming two-dimensional, with the longitudinal modal quantum number being frozen. 

Thermalization of the cavity photons is achieved by repeated absorption re-emission processes by the dye molecules, equilibrating the photon gas to the rovibrational temperature of the dye, which is at room temperature.
The dye molecules here fulfil the Kennard-Stepanov relation, a thermodynamic Boltzmann-type frequency scaling between absorption and emission spectra well known to apply for systems with both upper and lower electronic states rovibrational manifolds in thermal equilibrium \cite{Kennard, Stepanov, PhysRevA.54.4837}, as here achieved from frequent collisions with solvent molecules in the dye solution.
When the thermalization of the photon gas is faster than photon loss through e.g. the cavity mirrors, photons condense into the macroscopically occupied mode at the position of the low frequency cutoff on top of a spectrally broad thermal photon cloud.
In the opposite limit of photons leaving the cavity before they have a chance to thermalize with energetically excited states, corresponds to the usual laser case \cite{kirton2013nonequilibrium, schmitt2015thermalization}.

Other than in a usual laser in a photon Bose-Einstein condensation setup no inverted active medium is required to generate coherent optical emission, and spontaneous emission is retrapped \cite{klaersbose2010}. 
This makes the system particular attractive for ultraviolet or vacuum ultraviolet coherent optical sources, a spectral regime at which inversion, and correspondingly laser operation, is hard to achieve.
We are aware of ongoing other work aiming at high optical frequency coherent emission using quantum interference effects, forming "lasers without inversion" \cite{zibrov1995experimental, kocharovskaya1992amplification}.
A possible suitable medium for photon thermalization in the vacuum ultraviolet regime could be high pressure noble gas samples, which exhibit electronic transitions starting from the electronic ground state in this spectral regime, or both mixtures between atoms or molecules and noble gases.
Alkali-noble gas collisions are long known to be extremely elastic \cite{speller1979quenching, vogl2008spectroscopy}, and for the case of rubidium-argon gas mixtures at a few hundred bar of argon pressure the applicability of the thermodynamic Kennard-Stepanov scaling between absorption and emission has been validated for the visible rubidium D-lines \cite{moroshkin2014kennard}.
The latter results can be well described by a simple model that assumes equilibrium between external degrees of freedom and the internal structure of alkali-noble gas quasi-molecules in the electronic ground and excited state manifolds, as due to the frequent collisions in the dense buffer gas system.
For the case of xenon noble gas, vacuum ultraviolet absorption spectra have in earlier work been carried out up to \SI{20}{\bar} pressure \cite{borovich1973pressure}.
For simple molecules a large collection of both absorption and partly also emission data exists in the vacuum ultraviolet spectral regime, but mostly only for pure samples or at moderate buffer gas pressures \cite{lu2004analysis, zelikoff1953absorption, orkin1997rate}.

As a first step towards a photon condensate in the vacuum ultraviolet spectral regime, we have started experiments to test for the applicability of the Kennard-Stepanov relation of gas samples, to provide a spectroscopic test of possible photon gas thermalization.
Using the emission of a VUV deuterium lamp, we have recorded absorption spectra of xenon gas in the \SIrange{142}{160}{\nano \meter} wavelength range for up to \SI{30}{\bar} of pressure.
Moreover, spectra of ethylene molecules have been recorded in the \SIrange{155}{180}{\nano \meter} wavelength range for up to \SI{130}{\bar} of argon buffer gas pressure.

A further motivation for the here described work stems from the possibility to carry out collisional redistribution laser cooling of the dense gas samples.
In earlier work of our group, collisional laser cooling has been carried out with rubidium gases subject to near \SI{200}{\bar} argon buffer gas pressure, where frequent rubidium-argon collisions shift the alkali atomic transition into resonance with a far red detuned laser beam, while spontaneous decay occurs near the unperturbed resonance frequency \cite{vogl2009laser, yeh1979theory}.
The reaching of a sufficient alkali vapour pressure near \SI{1}{\milli \bar} here requires for a pre-heating of the sample cells, while gases that are gaseous at room temperatures, as ethylene, CO$_2$ or pure noble gases, could allow for redistribution laser cooling starting from room temperature conditions. 
The electronic transitions of such gases frequently are in the ultraviolet or vacuum ultraviolet spectral regime, as investigated in the here described work.

In the following, Section \ref{sec:Exp} describes the used experimental setup, and Section \ref{sec:xenon} gives obtained spectroscopic data for the case of xenon gas.
Further, in Section \ref{sec:ethylene} we present results obtained with molecular ethylene gas both for the case without and with noble gas buffer gas. 
Finally, Section \ref{sec:outlook} gives conclusions.

\begin{figure*}[t]
\begin{center}
\resizebox{.98\textwidth}{!}{%
 \includegraphics{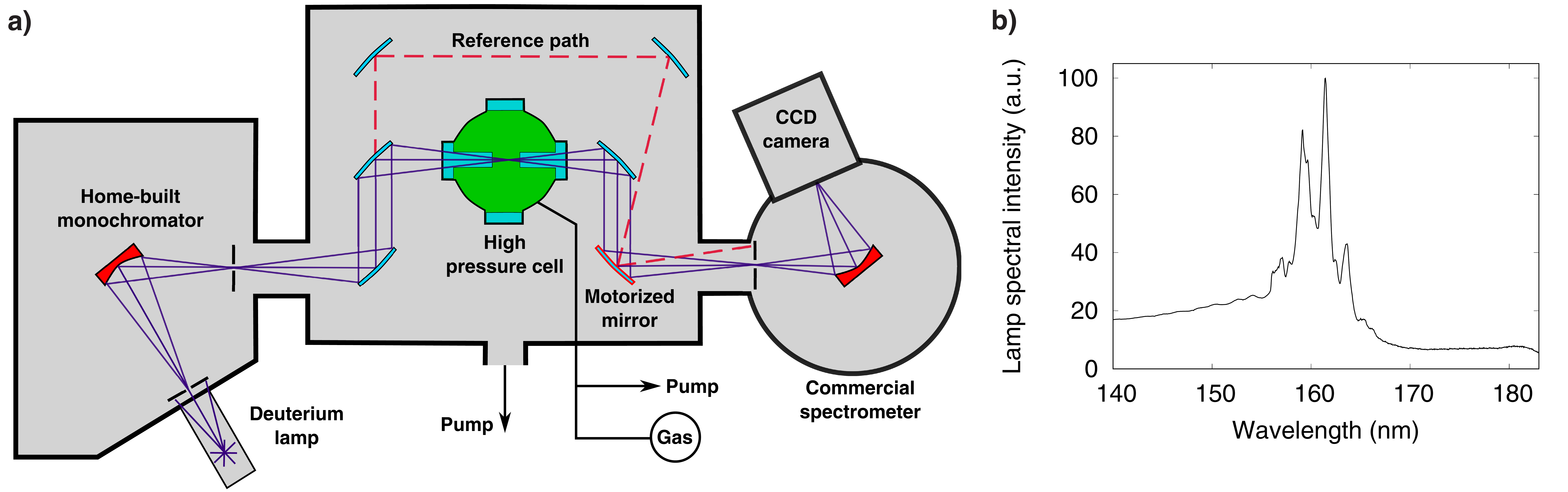}
}
\end{center}
\caption{\textbf{a)} Schematic representation of the experimental setup. Light emitted from the deuterium lamp is spectrally filtered by the first \mbox{monochromator} and send through a high pressure cell containing the gas to be analysed. A fraction of the light passes around the cell and is used to monitor the source output. A flip mirror is used to guide light from one of the two paths into the detection spectrometer.
All optical elements in this setup are coated with aluminium and a protective magnesium difluoride (MgF$_2$) layer. 
\textbf{b)} Deuterium source spectral intensity between \SIrange{155}{185}{\nano \meter}.}
\label{fig:Setup}       
\end{figure*}

\section{Experimental setup}
\label{sec:Exp}
\noindent
A schematic of the used experimental apparatus for VUV absorption measurements is shown in Figure \ref{fig:Setup}a. 
The optical setup including the high pressure cell containing the investigated gas sample is placed in a vacuum chamber evacuated to \SI{5e-5}{\milli \bar}.
As a light source for the spectroscopy setup we use a \SI{200}{\watt} deuterium lamp (model: Heraeus D200VUV), whose optical emission spans the spectral range from \SIrange{120}{400}{\nano \meter}. 
Its emission is not uniform within this spectral range, but rather exhibits variations by two orders of magnitude, see Fig. \ref{fig:Setup}b.
Therefore a two monochromator setup is used, which allows for a large dynamic range of the spectroscopy signal.
At first, a home-built grating monochromator (see Fig. \ref{fig:Setup}a) provides initial wavelength selection of the source, which besides the enhancement of the possible dynamic range of the spectroscopy setup also suppresses transitions in the gas sample driven by radiation with wavelength outside of the detection window.
The home-built monochromator is equipped with a grating with $1200$ groves/\si{\milli \meter} with a blaze angle optimized for \SI{200}{\nano \meter} radiation.
The concave grating focuses the emission to a first exit slit of \SI{400}{\micro \meter} width.
The width of spectral selection of this first monochromator is \SI{2.2}{\nano \meter}.
The spectrally filtered radiation is split into a reference and a measurement beam. 
The latter is focused into a high pressure cell containing the sample and then directed onto a commercial grating spectrometer (model: McPherson 234/302) consisting of a \SI{10}{\micro \meter} wide entrance slit and a concave grating with $2400$ lines/\si{\milli \meter}, which images the entrance slit onto an open nose VUV sensitive CCD camera.
The spectral resolution of the commercial spectrometer is \SI{0.05}{\nano \meter}.
The reference beam is guided around the sample cell and by means of a motorized mirror can be alternatively, to the measurement beam, sent onto the commercial spectrometer.
This allows for a normalization of the detected spectrum.
This was done for each data point individually since we found the intensity of the detected signal to fluctuate over time by nearly \SI{30}{\percent} on a \SI{30}{\minute} time-scale, an issue that is most likely due to polarization variations of the source which lead to intensity changes by the used optical elements.

The samples investigated in this work include high-purity ethylene, argon, helium, and xenon gases. 
High gas pressures, above \SI{40}{\bar}, are measured using a \SI{200}{\bar} gauge with a \SI{1}{\bar} resolution.
Intermediate pressures in the range between \SI{150}{\milli \bar} to \SI{40}{\bar} are determined using a digital gauge with a \SI{10}{\milli \bar} resolution.
For pressures below \SI{150}{\milli \bar} a fine needle gauge is utilized, giving a resolution of \SI{0.05}{\milli \bar}. 
The pressure cell is constructed to house pressures of up to \SI{200}{\bar}.
To ensure a largely contamination free environment the high pressure cell is repeatedly flushed with argon and evacuated to \SI{e-5}{\milli \bar}.
Before each absorption measurement a reference is taken to ensure that no residual absorption is observed.
In order to seal the cell MgF$_2$ step windows of different step lengths are used, which further allow to vary the absorption length in the cell.

To measure broadband absorption spectra both monochromators are first coupled, set to the same wavelength and subsequently scanned over the whole region of interest.
Since the width of the spectral selection of the first monochromator is \SI{2.2}{\nano\meter}, wavelength steps of \SI{1}{\nano\meter} are performed while measuring the absorption and lamp output for each wavelength setting.
The resulting signals are then, after being corrected for the current power of the deuterium lamp, used to create a spectrum spanning the region of interest.
Broadband absorption spectra are then determined from the Lambert-Beer law.

\begin{figure}[t]
\resizebox{0.6\textwidth}{!}{%
 \includegraphics{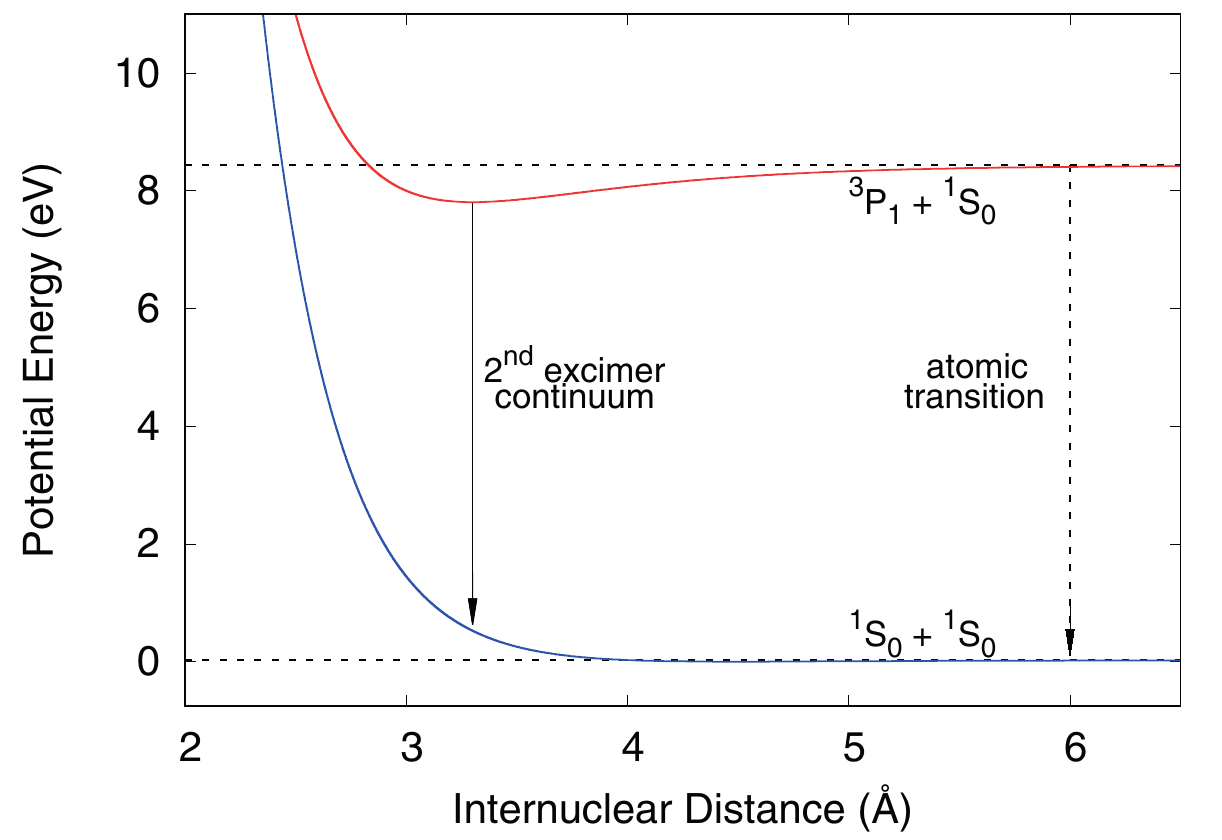}
}
\caption{Calculated energy in the xenon dimer system versus the internuclear distance \cite{dutuit1980synchrotron,docken1973spectroscopic}. The horizontal dashed lines give the asymptotic energy levels for infinitely large internuclear distance.}
\label{fig:Xenon_pot}       
\end{figure}

\begin{figure}[t]
\resizebox{0.6\textwidth}{!}{%
 \includegraphics{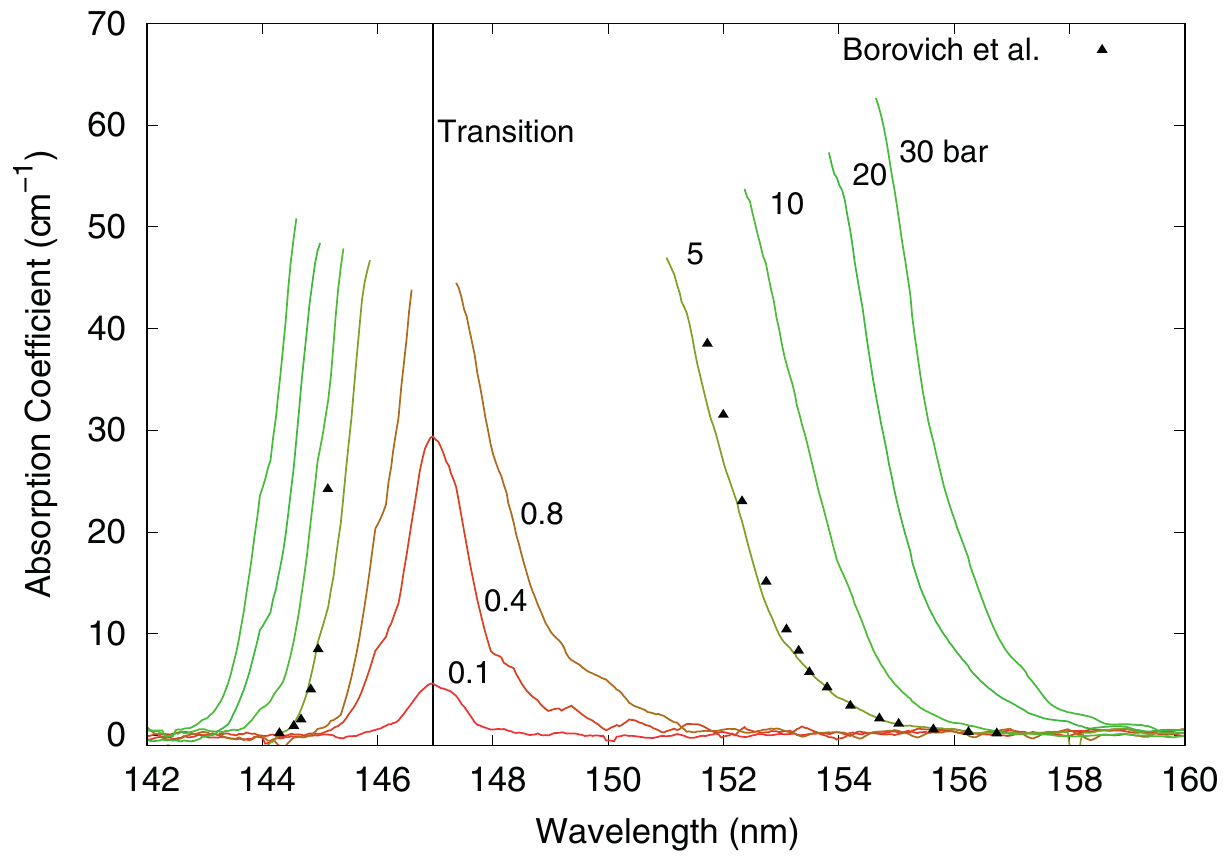}
}
\caption{Absorption coefficients of pure xenon in the wavelength range of \SIrange{142}{160}{\nano \meter} for pressures between \SIrange{0.1}{30}{\bar}. Xenon pressures are indicated next to the respective absorption spectrum. Saturated regions are omitted for clarity and the black triangles indicates comparable measurements from Borovich et al. for the \SI{5}{\bar} dataset \cite{borovich1973pressure}.}
\label{fig:Xenon}      
\end{figure}

\section{Xenon VUV absorption measurements}
\label{sec:xenon}
\noindent
To begin with, we have recorded xenon absorption spectra around the transition from the $5p^6$ ground state to the $5p^5 6s$ electronically excited state near \SI{147}{\nano\meter}.
At the used high xenon gas pressures frequent collisions between atoms occur and the quasi molecular character in a regime beyond the impact limit determines the line shapes.
Figure \ref{fig:Xenon_pot} shows calculated potential curves for the binary xenon system including the relevant levels. 
The lines in Figure \ref{fig:Xenon} show experimentally observed absorption data for various xenon pressures in the range of \SIrange{0.1}{30}{\bar}.
We observe significant pressure broadening, with the red wing being strongly enhanced with respect to the blue wing.
This is well understood from the potential curves, with the ground state for not to small internuclear distances being mostly flat and the excited state decreasing in energy.
Above a pressure of \SI{0.4}{\bar}, the absorption coefficient at the line center cannot be resolved due to the here very high optical density.
Correspondingly, we also cannot determine the linewidth for the high pressure data, but already at the quoted pressure value of \SI{0.4}{\bar} the linewidth exceeds the thermal energy $k_\text{B}T$ in wavelength units of \SI{0.5}{\nano\meter}  at room temperature for a wavelength of \SI{147}{\nano \meter}.
Our results are in good agreement with earlier measurements by Borovich et al., see the exemplary data points for a pressure of \SI{5}{\bar} in Figure \ref{fig:Xenon} \cite{borovich1973pressure}.

The spontaneous emission of the xenon system at pressures above \SI{50}{\milli \bar}, where the collision rate is sufficient that excimer molecules form faster than spontaneous decay, is known to be dominated by the second continuum around \SI{172}{\nano \meter}, see the corresponding potential curve of Fig. \ref{fig:Xenon_pot}.
Though a linear extrapolation of our presented data with pressure does not indicate a significant overlap with emission at the second continuum at pressures below values reachable with our cell (\SI{200}{\bar}), upon reaching of the critical pressure of \SI{58.4}{\bar} in the xenon system a highly nonlinear increase of the density with pressure does occur, which will strongly influence the spectra.
This has been indicated in earlier narrowband optical measurements \cite{PhysRevA.9.768}.
We are currently working on corresponding broadband measurements in this regime.

\section{Absorption spectroscopy of ethylene}
\label{sec:ethylene}

\begin{figure*}[t]
\resizebox{0.98\textwidth}{!}{%
 \includegraphics{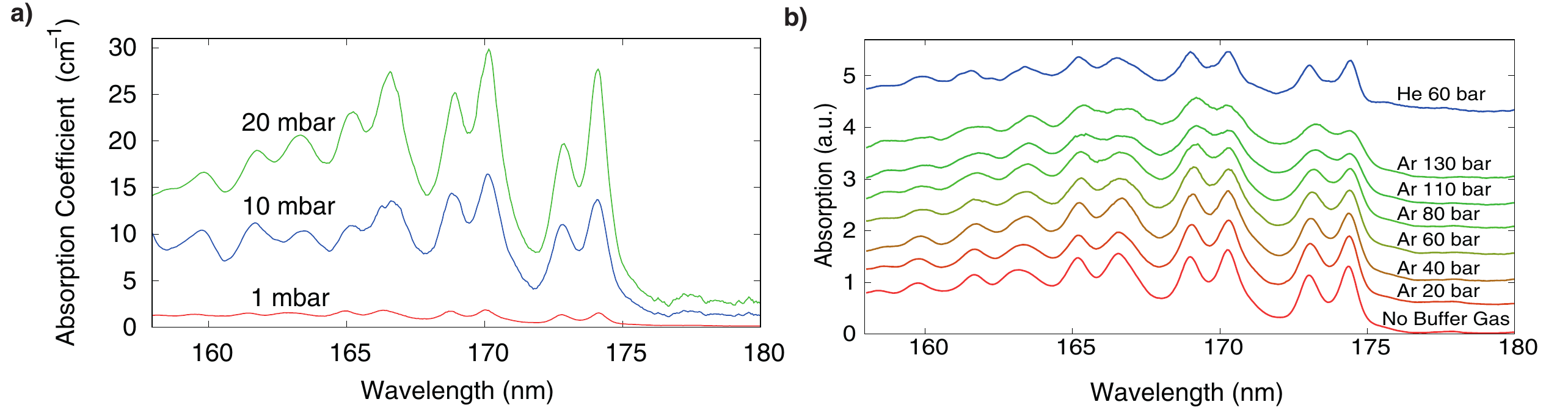}
 }
\caption{\textbf{a)} Absorption coefficients in the wavelength range of \SIrange{159.5}{177}{\nano \meter} for three different pressures of pure ethylene, \SI{1}{\milli \bar} $\left(\textit{red}\right)$, \SI{10}{\milli \bar} $\left(\textit{blue}\right)$, and \SI{20}{\milli \bar} $\left(\textit{green}\right)$. \textbf{b)} Absorption spectra for \SI{1}{\milli \bar} of ethylene with the addition of \SIrange{20}{130}{\bar} Argon. Spectra are vertically shifted by 0.5 for clarity. It is observed that the peaks in the ethylene spectrum experience a red-shift dependent on argon pressure. Further, a strong decrease in contrast between peak and underlying continuum is apparent. The top spectrum (blue line) gives an exemplary spectrum for helium buffer gas at \SI{60}{\bar} pressure.}
\label{fig:Ethylene-Ar}      
\end{figure*}

\noindent
An alternative promising approach for a thermalization medium for a photon Bose-Einstein condensate in the VUV spectral regime are simple molecules in high pressure noble buffer gas environments.
Ethylene molecules show electronic transitions in the relevant wavelength range, and the low pressure regime (below \SI{0.3}{\milli \bar}) of pure ethylene has previously been studied \cite{lu2004analysis, zelikoff1953absorption, orkin1997rate}.
To begin with, we have investigated corresponding spectra of pure ethylene up to \SI{20}{\milli \bar} pressure in the wavelength range between \SIrange{155}{180}{\nano \meter}, see Fig. \ref{fig:Ethylene-Ar}a.
The absorption coefficient scales proportionally with increasing ethylene concentration as expected from Beer's law, while no evident pressure-broadening effect can be observed.
Earlier experiments are in very good agreement with the present work, presenting comparatively negligible discrepancies (on the order of \SI{0.1}{\nano \meter}) in the observed peak positions.
Fig. \ref{fig:Ethylene-Ar}b gives corresponding ethylene spectra both without (bottom) and with argon buffer gas, for buffer gas pressures of up to \SI{130}{\bar}.
The ethylene molecular partial pressure here was \SI{1}{\milli \bar}.
One observes that the addition of the buffer gas broadens the individual resonances, while the overall spectral structure is not significantly modified.
Additionally we observe for the 3 visible doublets with minima at \SI{165.6}{\nano \meter}, \SI{168.4}{\nano \meter}, and \SI{173.5}{\nano \meter}, respectively, that the long wavelength peak is dominant at low buffer gas pressures compared to the short wavelength peak, which is inverted at high buffer gas pressures.
When using helium instead of argon as a buffer gas no significant modification of the spectra is observed, see the exemplary curve in Fig. \ref{fig:Ethylene-Ar}b.

\begin{figure}[t]
\resizebox{.6\textwidth}{!}{%
 \includegraphics{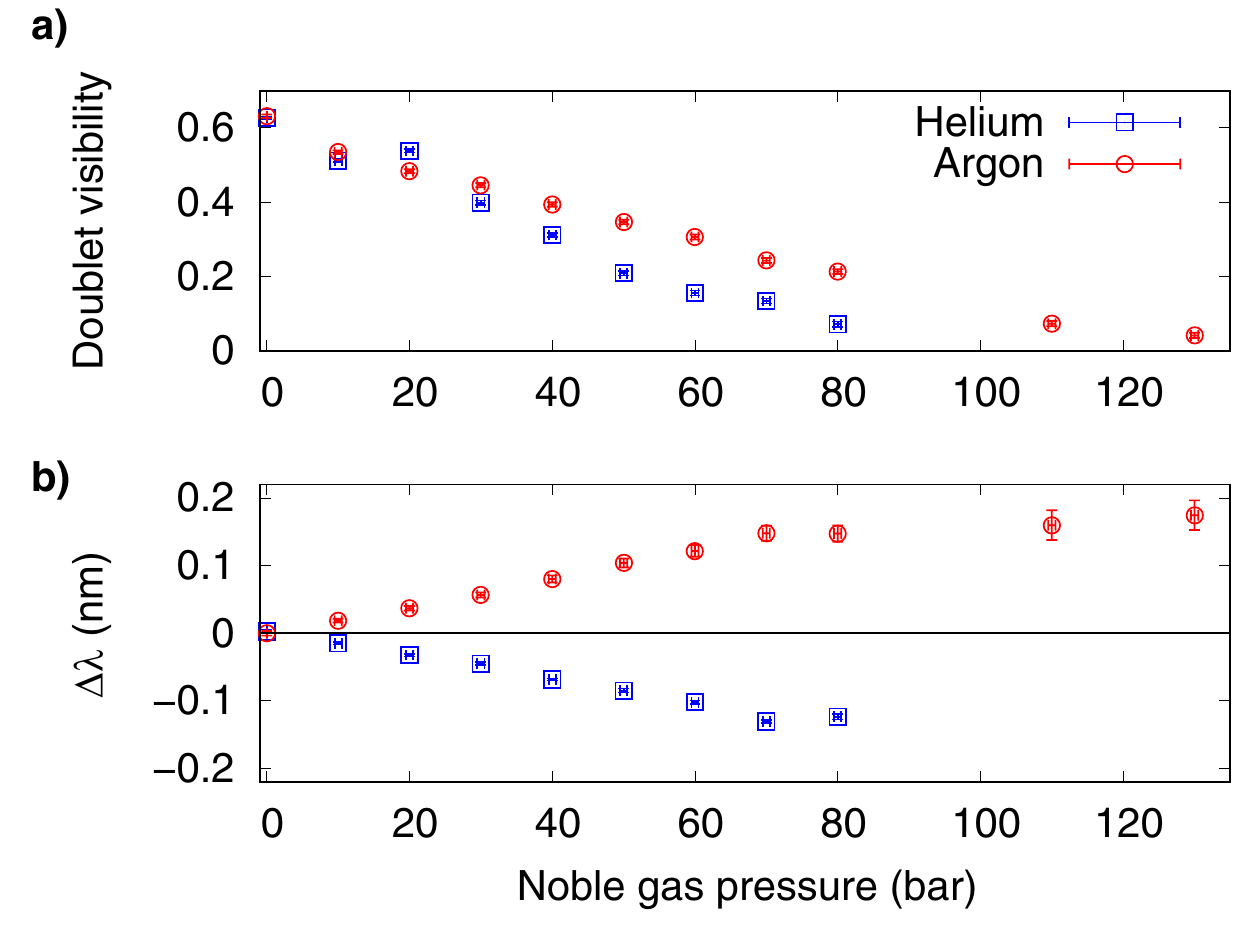}
}
\caption{\textbf{a)} Extracted visibility doublet (see text) and \textbf{b)} peak position for the peak found at \SI{174.1}{\nano \meter} in \SI{1}{\milli \bar} of ethylene with added argon $\left(\textit{red cirlces}\right)$ and helium $\left(\textit{blue squares}\right)$ buffer gas versus the noble gas pressure. While both noble gases lead to a similar change in visibility, an opposing effect can be seen with respect to the wavelength shift. }
\label{fig:Ethylene-Shift}       
\end{figure}

To quantify the pressure broadening by the buffer gas we focus on the spectral peak near \SI{174.1}{\nano \meter}.
As we cannot unambiguously  determine the width of this line due to the many nearby resonances and the asymmetry of the line shapes, we instead use a normalized doublet visibility $V=\left(\kappa_{max}-\kappa_{min}\right)/\left(\kappa_{max}+\kappa_{min}\right)$, where $\kappa_{max}$ and $\kappa_{min}$ denote the absorption coefficients on resonance and at the minimum between the resonances of the doublet structure at \SI{172.8}{\nano \meter} and \SI{174.1}{\nano \meter} respectively, to quantify the pressure broadening, see also Fig. \ref{fig:Ethylene-Ar}b.
The variation of the doublet visibility for both argon and helium buffer gas with pressure is shown in Fig. \ref{fig:Ethylene-Shift}a. 
The data shows that the loss of doublet visibility with pressure and correspondingly also the pressure broadening is somewhat larger for the case of helium than for argon buffer gas. 
Fig. \ref{fig:Ethylene-Shift}b shows the pressure shift of the \SI{174.1}{\nano \meter} resonance for both of the buffer gases.
We observe a pressure shift of similar magnitude, but of opposite sign for the two noble gases.
For pressures of up to \SI{60}{\bar} the shift is linear to good approximation, with a slope of \SI{0.1}{\nano\meter} per \SI{50}{\bar}, while for higher pressures a saturation of the shift is observed.
At \SI{130}{\bar} argon buffer gas pressure, the observed pressure shift is \SI{0.18}{\nano \meter}, corresponding to $0.29$\,$k_\text{B}T$ in wavelength units at room temperature  for a wavelength of \SI{174}{\nano \meter}.
This shift is roughly a factor of 8 smaller than the pressure broadening responsible for the above discussed reduction of the visibility of the doublet spaced by \SI{1.4}{\nano \meter}.

\section{Conclusions and outlook}
\label{sec:outlook}
\noindent
To conclude, we have presented absorption spectra of both xenon gas and binary mixtures of ethylene with noble gases at high pressure.
Our measurements show a broadening of resonances of order of the thermal energy $k_\text{B}T$.
For the case of the ethylene data, both the data with and without noble buffer gas can provide valuable input for future calculations of molecular potentials.

In the future, we plan to test for the thermodynamic Kennard-Stepanov scaling of the dense gas samples, for which additionally fluorescence spectra at the corresponding parameters are required.
Besides the here presented xenon system, the lighter noble gases offer even shorter wavelength closed electronic transitions, as candidate systems for photon condensates in the deeper ultraviolet spectral regime.
Provided that the Kennard-Stepanov relation in the dense system holds, this implies a high quantum efficiency of the involved transitions \cite{klaers2014thermalization}.
Besides photon Bose-Einstein condensation, this would make the system also a promising candidate for collisional redistribution laser cooling starting from room temperature \cite{vogl2009laser}.

We acknowledge support of the Deutsche Forschungsgemeinschaft (within SFB/TRR 185 and We1748-17) and the European Research Council (INPEC).
We thank G. Wallstabe for his contributions in the early phase of this experiment.  

\bibliographystyle{spphys}
\bibliography{Paper_refs}
\end{document}